\documentclass[aps,preprint,a4paper,showpacs,amsmath,amssymb,floatfix]{revtex4}
\usepackage{graphicx}
\usepackage{dcolumn}
\usepackage{bm}
\usepackage{natbib}
\newcommand{\Op}[1]{{\boldsymbol{\mathrm{\hat{#1}}}}}

\newcommand{\beq}{\begin{equation}}
\newcommand{\eeq}{\end{equation}}
\newcommand{\beqar}{\begin{eqnarray}}
\newcommand{\eeqar}{\end{eqnarray}}
\newcommand{\bea}{\begin{eqnarray}}
\newcommand{\eea}{\end{eqnarray}}
\newcommand{\bcen}{\begin{center}}
\newcommand{\ecen}{\end{center}}

\newcommand{\half}{\frac{1}{2}}
 
\begin{document}

\title{On the Local and Global Approaches to Quantum Transport  and Violation of the Second-law of Thermodynamics\\
}

\author{Amikam Levy and Ronnie Kosloff}

\affiliation{
Institute  of Chemistry,
The Hebrew University of Jerusalem, Jerusalem 91904, Israel\\
}

\begin{abstract}
Clausius statement of the second law of thermodynamics reads: Heat will flow spontaneously from a hot to cold reservoir.
This statement should hold for transport of energy through a  quantum network composed of small subsystems each coupled 
to a heat reservoir. When the coupling between nodes is small, it seems reasonable to construct a local master equation 
for each node in contact with the  local reservoir. The energy  transport through the network
is evaluated by calculating the energy flux after the individual nodes are coupled. We show by analyzing the 
most simple network composed of two quantum nodes coupled to a hot and cold reservoir, that the local description
can result  in heat flowing from cold to hot reservoirs, even in the limit of vanishing coupling between the nodes.
A global derivation of the master equation which prediagonalizes the total network Hamiltonian and within this framework derives
the master equation, is always consistent with the second-law of thermodynamics.
\end{abstract}

\maketitle
\section{Introduction} 
\label{sec:introduction}
Transport of energy in and out of a quantum device is a key issue in emerging technologies.
Examples include molecular electronics, photo-voltaic devices, quantum refrigerators and quantum heat engines  \cite{vanderwiel03,Kohler05,levy14}.
A quantum network composed of quantum nodes each coupled to local reservoir and to other nodes constitutes the network.
The framework for describing such devices is the theory of open quantum systems. 
The dynamics is postulated employing completely positive quantum master equations \cite{lindblad76,gorini276}.
Solving the dynamics allows to calculate the steady state transport of energy through the network.
\par
It is desirable to have the framework consistent with thermodynamics. 
The first law of thermodynamics is a conservation law of energy;
the energy of an isolated system is constant and can be divided into heat and work \cite{alicki79}.
The dynamical version of the second law of thermodynamics states that for an isolated system the rate of entropy production is non-negative \cite{k281}. 
For a typical quantum device the second law can be expressed as,
\begin{equation}
\frac{d}{dt} \Delta {\cal S}^u = \frac{d {\cal S}_{int}}{dt} +\frac{d {\cal S}_{m}}{dt}-\sum_i\frac{{\cal J}_i}{T_i} \geq 0,
\end{equation} 
where $\dot {\cal S}_{int}$ is the rate of entropy production due to internal processes, expressed by the von Neumann entropy.
$\dot {\cal S}_m$ is the entropy flow associated with matter entering the system, and the last term is the contribution 
of heat flux, ${\cal J}_i$, from the reservoir $i$. 
\par
Microscopic derivation of a global Markovian master equation (MME) of Linblad-Gorini-Kossakowski-Sudarshan (LGKS) form \cite{lindblad76,gorini276}, for the network is usually intricate. 
The local approach simplifies this task \cite{mari2012,linden10,restrepo14,atalaya12,zoller04,breuer07,brunner14}. It is commonly considered that if the different parts of the network are weakly coupled to each other, a local master equation is sufficient to describe all the properties of the network.
We will show that the local approach is only valid for local observables such as the population of each node, and is not valid for non-local observables describing energy fluxes.  
\section{The Network Model}
\begin{figure}[htbp]
\center{\includegraphics[scale=0.5]{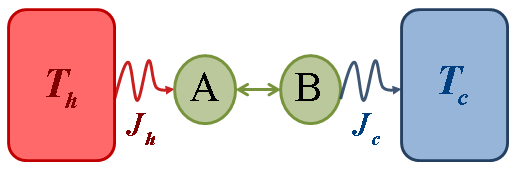}}
\caption{The heat transfer network model; heat is transferred from a hot bath at temperature $T_h$ to the a colder bath at temperature $T_c$.
The heat current is mediated by two coupled subsystems $A$ and $B$, where subsystem $A$ is connected to the hot bath and subsystem $B$ is connected to the cold bath.}
\label{fig:A}
\end{figure}

The simplest network model composed of two nodes shown in fig. \ref{fig:A} and 
is sufficient to demonstrate the distinction between the local and global approach.
Heat is transported between two subsystems $A$ and $B$, where each is coupled to a single heat bath with temperature $T_h$ and $T_c$. 
The two subsystems are weakly coupled to each other.  
The global Hamiltonian is of the form:
\begin{equation}
\Op  H = \Op H_A + \Op H_B + \Op H_{AB} + \Op H_h +\Op H_c + \Op H_{Ah} +\Op H_{Bc}.
\end{equation}
The bare network Hamiltonian, is  $\Op H_0= \Op H_A + \Op H_B $ where the node Hamiltonians are 
$\Op H_A = \omega_h \Op a^{\dagger} \Op a$ 
and $\Op H_B = \omega_c \Op b^{\dagger} \Op b$, which are composed of either two harmonic oscillators (HO) or of two two-level systems (TLS), 
depending on the commutation relation.
\begin{equation}
\Op a \Op a^{\dagger} +\delta \Op a^{\dagger} \Op a =1~ , \Op a \Op a +\delta \Op a \Op a=0 ~,\Op b \Op b^{\dagger} +\delta \Op b^{\dagger} \Op b =1 , ~\Op b \Op b +\delta \Op b \Op b=0
\label{11}
\end{equation}
with $\delta = 1$ for the TLS and $\delta = -1$  for oscillators.  
The interaction between the system $A$ and $B$ is described by the swap Hamiltonian, 
$\Op H_{AB} = \epsilon (\Op a^{\dagger}\Op b + \Op a \Op b^{\dagger})$, with $\epsilon > 0$. 
The hot (cold) baths Hamiltonians are denoted $\Op H_{h(c)}$, where $T_h> T_c$. The system-bath interaction is given by, 
$\Op H_{Ah} = g_h(\Op a + \Op a^{\dagger})\otimes \Op R_h $ and $\Op H_{Bc}= g_c(\Op b + \Op b^{\dagger}) \otimes \Op R_c$, 
with $\Op R_{h(c)}$ operators belonging the hot (cold) bath Hilbert space, and $g_{h(c)}$ are the system-baths coupling parameter.
\par
The dynamics of the reduced system $A + B$ is governed by the Master equation,
\begin{equation}
 \frac{d}{dt}\Op \rho_s = -i[\Op H_0 +\Op H_{AB},\Op \rho_s] + {\cal L}_h \Op  \rho_s+ {\cal L}_c \Op \rho_s.
\end{equation}
With the LGKS dissipative terms, ${\cal L}_{h(c)}$, which differ for the local and global approaches.
At steady state the heat flow from the hot (cold) bath is given by,  
\begin{equation}
 {\cal J}_{h(c)}=Tr[({\cal L}_{h(c)}\Op  \rho_s)(\Op H_0 + \Op H_{AB})],
\end{equation}
where $\Op \rho_s$ is the steady state density operator.
\section{Local Approach}
\begin{figure}[htbp]
\center{\includegraphics[scale=0.35]{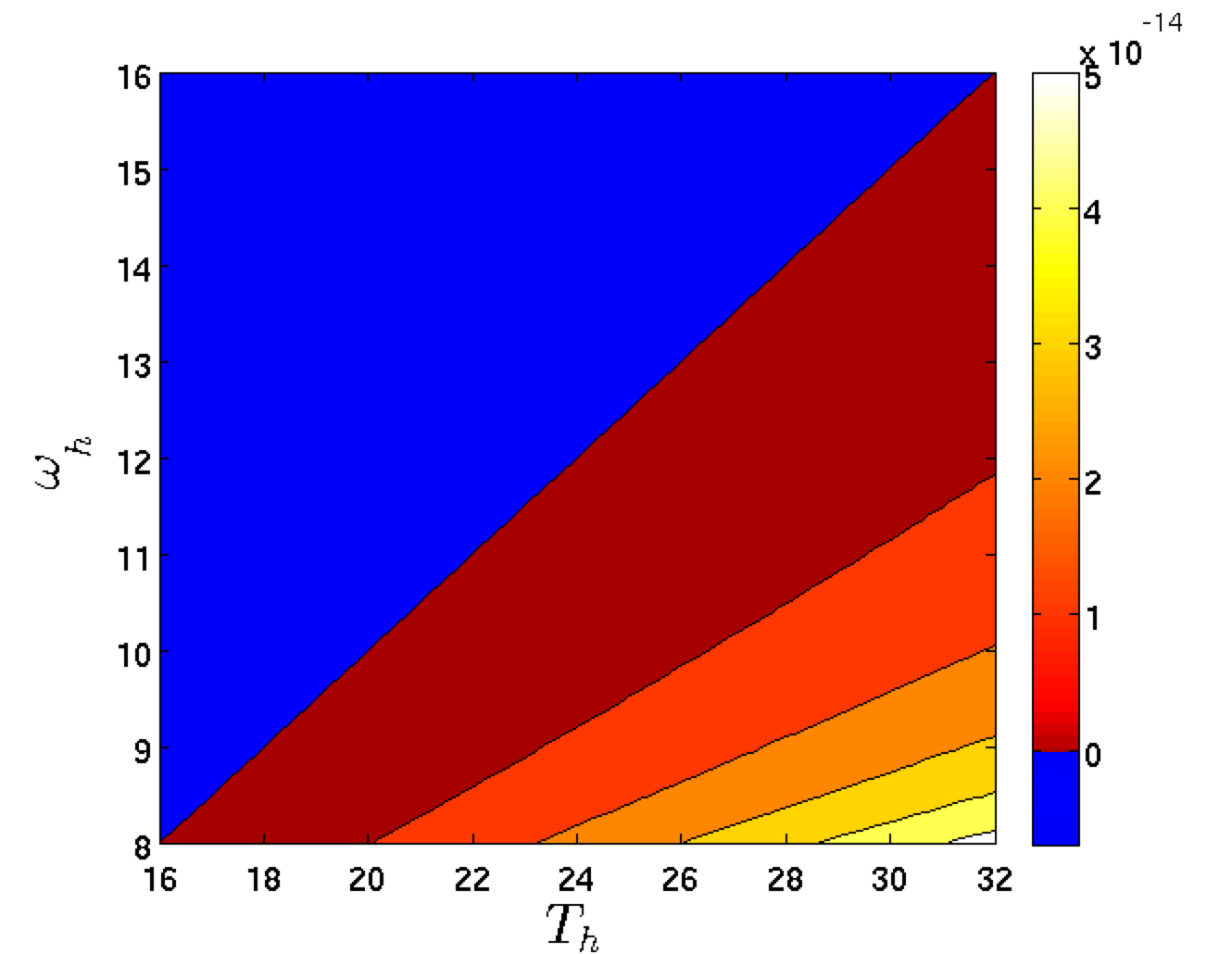}}
\caption{The rate of entropy production $\Delta {\cal S} ^u$ in the local description, as function of $\omega_h$ and $T_h$.
The blue area correspond to negative entropy production rate, a clear violation of the second law. 
The borderline between the blue and the red zones correspond to $\omega_h/T_h = \omega_c/T_c$.
Here $T_c=10$, $\omega_c = 5$, $\epsilon = 10^{-4}$ and $\kappa = 10^{-7}$.}
\label{fig:B}
\end{figure}

In the local approach it is assumed that the inter-system coupling does not affect the system bath coupling.
Therefore in the derivation of the MME the Hamiltonian $\Op H_{AB}$ is ignored and the dissipative terms takes the form,
\begin{equation}
\label{eq:lhl}
{\cal L}_h \Op \rho_s = \gamma_h \left( \Op a \Op \rho_s \Op a^{\dagger} -\half\{\Op a^{\dagger}\Op a,\Op \rho_s\} 
+ e^{-\beta_h \omega_h}(\Op a^{\dagger} \Op \rho_s \Op a -\half\{\Op a \Op a^{\dagger},\Op \rho_s\})\right),
\end{equation}
and
\begin{equation}
\label{eq:lcl}
{\cal L}_c \Op \rho_s = \gamma_c \left( \Op b \Op \rho_s \Op b^{\dagger} -\half\{\Op b^{\dagger} \Op b,\rho_s\} + e^{-\beta_c \omega_c}(\Op b^{\dagger} \Op \rho_s \Op  b -\half\{\Op b \Op b^{\dagger},\Op \rho_s\})\right).
\end{equation}
when the node to node coupling is zero, $\Op H_{AB}=0$, each of the local master equations eq. (\ref{eq:lhl}) and eq. (\ref{eq:lcl})
drives the local node to thermal equilibrium.
The dynamics of the network is completely characterized by the expectation values of four operators: Two local observables
$\langle \Op a^{\dagger} \Op a \rangle, \langle \Op b^{\dagger} \Op b \rangle$, and two $AB$ correlations 
$\langle \Op X \rangle \equiv \langle \Op a ^{\dagger}\Op b + \Op a \Op b^{\dagger} \rangle $ and 
$\langle \Op Y\rangle \equiv i\langle \Op a^{\dagger} \Op b - \Op a \Op b^{\dagger} \rangle $ with 
$ \langle ~\cdot~\rangle  \equiv tr\{\Op \rho_s \cdot\}$.  For the dynamics we obtain: 
\begin{eqnarray}
\begin{array}{ll}
\label{eq:equation of motion}
\frac{d}{dt}\langle \Op a^{\dagger}\Op a\rangle = -\gamma_h(1+\delta e^{-\beta_h \omega_h}) \langle \Op a^{\dagger}\Op a\rangle+ \gamma_h e^{-\beta_h \omega_h} - \epsilon \langle \Op Y\rangle \\
\frac{d}{dt}\langle \Op b^{\dagger}\Op b\rangle = -\gamma_c(1+\delta e^{-\beta_c \omega_c}) \langle \Op b^{\dagger}\Op b\rangle+ \gamma_c e^{-\beta_c \omega_c} + \epsilon \langle \Op Y\rangle \\
\frac{d}{dt}\langle \Op X\rangle = -\half \left( \gamma_h(1+\delta e^{-\beta_h \omega_h})+\gamma_c(1+\delta e^{-\beta_c \omega_c})\right) \langle  \Op X\rangle  +(\omega_h -\omega_c) \langle \Op Y\rangle \\
\frac{d}{dt}\langle \Op Y\rangle = -\half \left( \gamma_h(1+\delta e^{-\beta_h \omega_h})+\gamma_c(1+\delta e^{-\beta_c \omega_c})\right) \langle \Op Y\rangle -(\omega_h -\omega_c) \langle \Op X\rangle +2\epsilon(\langle \Op a^{\dagger}\Op a\rangle -\langle \Op b^{\dagger}\Op b \rangle)\\
\end{array} 
\end{eqnarray} 
The rate $\gamma >0$ depends on the the specific properties of the bath and its interaction with the system.
Equations (\ref{eq:equation of motion})  fulfill  the dynamical version of the first law of thermodynamics: The sum of all energy (heat) currents at steady state is zero, ${\cal J}_h + {\cal J}_c = 0$.
The heat flow from the hot heat bath can be cast in the form (see Appendix for details).
\begin{equation}
\label{eq:jh}
 {\cal J}_h = (e^{\beta_c\omega_c} - e^{\beta_h \omega_h}){\cal F},
\end{equation}
where ${\cal F}$ is a function of all the parameters of the system, which is always positive, and is different for the HO and TLS medium.
The Clausius statement for the second law of thermodynamics implies that heat can not flow from a cold body to a hot body without external work being performed on the system.
It is apparent from eq.(\ref{eq:jh}), that the direction of heat flow depends on the choice of parameters.
For $\frac{\omega_c}{T_c} < \frac{\omega_h}{T_h}$ heat will flow from the cold bath to the hot bath, thus the second law is violated even at vanishing small $AB$ coupling, Cf. fig. \ref{fig:B}.
The breakdown of the second law has been examined in several models, see \cite{novotny02} and references therein. 
In \cite{novotny02} a Fermionic transport model between two heat baths at the same temperature was studied in the weak system-bath coupling limit MME and was compared to a solution within the formalism of nonequilibrium Green functions.
At steady state, the current between the baths according to the weak coupling MME is nonzero, which implies a violation of the second law in the sense that heat flows constantly between two heat baths at the same temperature.
This sort of violation can also be observed in eq.(\ref{eq:jh}) when taking $T_h=T_c$. 
It was claimed in \cite{novotny02} that the violation of the second law is a consequence of neglecting higher-order coherent processes between the system and the baths due to the weak coupling limit. 
In fact, the treatment introduced in \cite{novotny02} corresponds to the local approach described above. 
Next, we introduce a proper weak coupling MME, which always obeys the second law of thermodynamics.
\section{Global Approach}
The global approach is based on the holistic perception where the MME
is derived in the eigen-space representation of the combined system $A+B$. 
The reduced system, $A+B$, is first diagonalized, then the new basis set is used to expand the system-bath interactions. 
Finally, the standard weak system-bath coupling procedure is introduced to derive the MME \cite{davies74,breuer}.
This approach accounts for  a shift in the spectrum of the subsystems $A$ and $B$ due to the coupling parameter $\epsilon$.
But more importantly, it creates an effective coupling of the system $A$ with the cold bath and of the system $B$ with the hot bath. 
This indirect coupling absent in the local approach is crucial, and essentially saves the second law of thermodynamics.
The global MME, by construction, obeys Spohn's inequality and therefore is consistent with the second law of thermodynamics \cite{spohn78}.
\par
In it's diagonal form the Hamiltonian $\Op H_0 + \Op H_{AB}$ is given by,
\begin{equation}
 \Op H_S = \omega_+ \Op d_{+}^{\dagger} \Op d_+ +\omega_- \Op d_{-}^{\dagger} \Op d_- .
\end{equation}
Where we have defined the operators $\Op d_+ = \Op a \cos(\theta) + \Op b\sin(\theta)$ and $\Op d_- = \Op b \cos(\theta) - \Op a \sin(\theta)$, with $\cos^2(\theta)= \frac{\omega_h -\omega_-}{\omega_+ - \omega_-}$ 
and $\omega_{\pm}= \frac{\omega_h +\omega_c}{2} \pm \sqrt{(\frac{\omega_h -\omega_c}{2})^2 + \epsilon^2}$. 
For Bosons, the commutation relations of the operators are preserved, i.e. $[\Op d_{\pm}, \Op d_{\pm}^{\dagger}] = 1$, where all other combinations are zero.
For TLS nodes the expressions are more intricate and therefore we restrict the analysis to the harmonic nodes.
Following the standard weak coupling limit, in the regime where $\omega_- > 0$ the dissipative terms of the MME reads,
\begin{eqnarray}
\begin{array}{ll}
\label{eq:lhg}
{\cal L}_h\Op \rho_s = \gamma_h^{+}\cos^2(\theta) \left( \Op d_+ \Op \rho_s \Op d_+^{\dagger} -\half\{\Op d_+^{\dagger}\Op d_+,\Op \rho_s\} + 
e^{-\beta_h \omega_+}(\Op d_+^{\dagger} \Op \rho_s \Op d_+ -\half\{\Op d_+ \Op d_+^{\dagger},\Op \rho_s\})\right) \\
~~~~~~~ + \gamma_h^{-}\sin^2(\theta) \left( \Op d_- \Op \rho_s \Op d_-^{\dagger} -\half\{\Op d_-^{\dagger} \Op d_-,\Op \rho_s\} 
+ e^{-\beta_h \omega_-}(\Op d_-^{\dagger} \Op \rho_s \Op d_- -\half\{\Op d_- \Op d_-^{\dagger},\Op \rho_s\})\right) 
\end{array} 
\end{eqnarray} 
and
\begin{eqnarray}
\begin{array}{ll}
\label{eq:lcg}
{\cal L}_c \Op \rho_s = \gamma_c^{+}\sin^2(\theta) \left( \Op d_+\Op \rho_s \Op d_+^{\dagger} -\half\{\Op d_+^{\dagger}\Op d_+,\Op \rho_s\} 
+ e^{-\beta_c \omega_+}(\Op d_+^{\dagger} \Op \rho_s \Op d_+ -\half\{\Op d_+ \Op d_+^{\dagger},\Op \rho_s\})\right) \\
~~~~~~~ + \gamma_c^{-}\cos^2(\theta) \left(\Op  d_- \Op \rho_s \Op d_-^{\dagger} -\half\{\Op d_-^{\dagger}\Op d_-,\Op \rho_s\} 
+ e^{-\beta_c \omega_-}(\Op d_-^{\dagger} \Op \rho_s \Op d_- -\half\{\Op d_-\Op d_-^{\dagger},\Op \rho_s\})\right) 
\end{array} 
\end{eqnarray}   
with $\gamma_{h(c)}^{\pm}=\gamma_{h(c)}(\omega_{\pm})$. 
The calculated steady state heat flow from the hot bath is given by,
\begin{equation}
\begin{array}{ll}
  {\cal J}_h = \frac{\left(e^{\beta _c \omega _-}-e^{\beta _h \omega _-}\right) \gamma _{c}^{-} \gamma _{h}^{-} \omega _-}{\sin^{-2}(\theta) e^{\beta _h \omega _-} \left(-1+e^{\beta _c \omega _-}\right) \gamma _{c}^- +e^{\beta _c \omega _+} \left(-1+e^{\beta _h \omega _-}\right) \cos^{-2}(\theta) \gamma _{h}^-}\\ 
~~~ +\frac{\left(e^{\beta _c \omega _+}-e^{\beta _h \omega _+}\right) \gamma _{c}^+ \gamma _{h}^+ \omega _+}{e^{\beta _h \omega _+} \left(-1+e^{\beta _c \omega _+}\right) \cos^{-2}(\theta) \gamma _{c}^+ +\sin^{-2}(\theta) e^{\beta _c \omega _+} \left(-1+e^{\beta _h \omega _+}\right) \gamma _{h}^+}
\end{array}
\end{equation}
which is positive for all physical choice of parameters.
Rewriting eq.(\ref{eq:lhg}) and eq.(\ref{eq:lcg})  in the local basis, the effective coupling of the of subsystem $A$ with the cold bath and of subsystem $B$ with the hot bath is immediately apparent (see Appendix for details).
These equations converge to eq.(\ref{eq:lhl}) and eq. (\ref{eq:lcl}) for $\epsilon=0$. 
\par
To further study the dynamics of $A$ and $B$, the explicit form of heat baths is specified, characterizing  the rates $\gamma$ \cite{k275}:
\begin{equation}
\gamma_l \equiv \gamma_l(\Omega) = \pi \sum_{k} |g_l(k)|^2 \delta (\omega(k) - \Omega)[1- e^{-\beta_l\omega(k)}]^{-1},
\end{equation}
where $\omega(k)$ are the frequencies of the baths modes. 
For the case of  a 3-dimensional phonon bath with a linear dispersion  relation  the relaxation rate  can be expressed as:
\begin{equation}
\gamma_l(\Omega) =  \kappa\Omega^3 [1- e^{-\beta_l\Omega}]^{-1},
\end{equation}
where $\kappa >0$ embodies  all the constants and is proportional to the square of the system-bath coupling.
\begin{figure}[htbp]
\center{\includegraphics[scale=0.4]{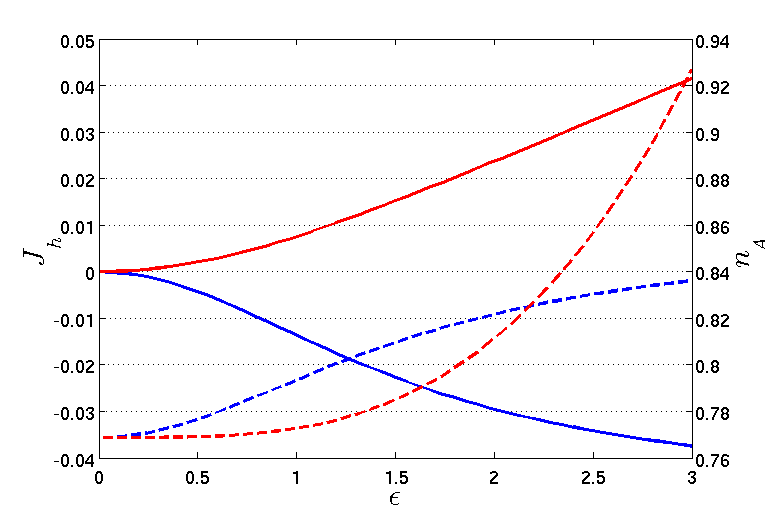}}
\caption{The heat current ${\cal J}_h$ and the population as function of the coupling parameter $\epsilon$
evaluated in the local (blue line) and the global (red line) approaches. 
The population of subsystem A (dashed line), and the heat flow from the hot bath ${\cal J}_h$ (solid line).
Here $T_h=12$, $T_c=10$, $\omega_h = 10$, $\omega_c = 5$ and $\kappa = 10^{-4} $.}
\label{fig:1}
\end{figure}
\par
The steady state observables of the local and global approached are compared in fig. \ref{fig:1} as a function of the node-to-node coupling strength $\epsilon$.
For local observables such as the local population $\Op n_A \equiv \left <\Op a^{\dagger}\Op a \right>$ 
the two approaches converge to the thermal population when 
$\epsilon \ll \{\omega_h,\omega_c, \sqrt{|\omega_h - \omega_c|}\}$.
However, the non-local observables such as the current ${\cal J}_h$ deviate  qualitatively. 
In the local approach when $\frac{\omega_c}{T_c} < \frac{\omega_h}{T_h}$ the second-law is violated:
the heat flow becomes negative for all values of the coupling $\epsilon$ while for the global approach
 ${\cal J}_h$ is always positive Cf. fig. \ref{fig:1}.
\begin{figure}[htbp]
\center{\includegraphics[scale=0.45]{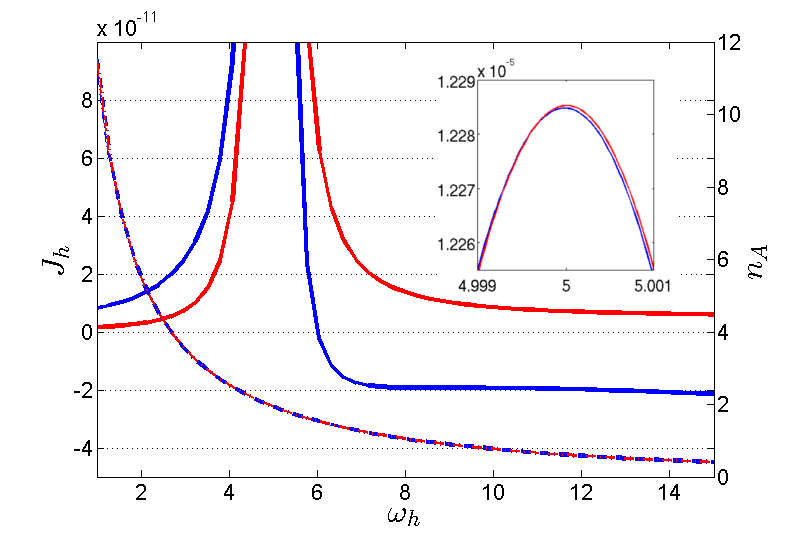}}
\caption{Comparison between the local (blue line) and the global (red line) approaches. 
The population of subsystem A (dashed line), and the heat flow from the hot bath ${\cal J}_h$ (solid line),  as function of $\omega_h$.
The inset describes the domain of near resonance $\omega_h \approx \omega_c$.
Here $T_h=12$, $T_c=10$, $\omega_c = 5$, $\epsilon = 10^{-3}$ and $\kappa = 10^{-7}$.}
\label{fig:2}
\end{figure}    
\par
The local approach is also not reliable even for parameters where the second-law 
is obeyed: $\frac{\omega_c}{T_c} > \frac{\omega_h}{T_h}$.
Deviations from the exact global approach appear
in the favorable domain of small $\epsilon$, as seen in  fig. \ref{fig:2} displaying ${\cal J}_h$ 
for a wide range of $\omega_h$.
It is noteworthy that the behavior of the heat flows observed in fig. \ref{fig:2} will be the same for all $\epsilon$, also when $\epsilon \ll \kappa$.
The only domain where the global approach breaks down do on resonance,wn is on resonance, when $\omega_h = \omega_c$ and $\epsilon < \kappa$. 
At this point, the secular approximation is not justified since the two Bohr frequencies $\omega_{\pm}$ are not well separated, and on the time scale $1/\kappa \omega^3 $, 
one can not neglect rotating terms such as $e^{i2\epsilon}$ \cite{plenio10}.    
\par
Additional insight is obtained when examining the covariance matrix for the two-mode Gaussian state (see Appendix for details).
The correlations between subsystems $A$ and $B$ is fully determined by the set of correlation functions $\{cor(x_A,x_B), cor(x_A,p_B), cor(p_A,x_B),\\ cor(p_A,p_B)\}$. Here $\{x,p\}$ are the position and momentum coordinates of the subsystems.  
In both approaches $cor(x_A,x_B)$ and $cor(p_A,p_B)$ are equal for small $\epsilon$. 
The two additional correlations, $cor(x_A,p_B)$ and $cor(p_A,x_B)$, vanish at steady state in the global approach, where in the local approach they remain finite. 
Thus, in the local approach the nodes are  over correlated compared to the global approach.
It should be noted that in steady state non of the approaches  generate entanglement. 
The two-mode Gaussian state is a separable state according to the separability criterion for continuous variable systems \cite{simon99,zoller99}, 
\par
To summarize: As expected, the local dynamical approach  is incorrect for strong coupling between the subsystems.
In the weak coupling limit, local observables converge to their correct value.
The non-local observables such as heat currents are qualitatively and quantitatively erroneous in the local MME.
A strong indication is the  violation of the second law of thermodynamics.
The completely positive LGKS generator is a desired form for the master equation. 
However, for consistency with the physical world, a microscopic global derivation of the master equation is required.
Such approaches are consistent with thermodynamics \cite{palao13,paz,k114,gelbwaser13,kolar13}.
\par
\acknowledgments
We want to thank Robert Alicki, Lajos Diosi and Angel Rivas for fruitful discussions and helpful comments.
This work was supported by the Israel Science Foundation and by the COST action MP1209 "Thermodynamics in the quantum regime".

\section{Appendix}
\subsection{Local Approach Heat Flow} 
\label{sec:Heat flow}
The heat flow from the hot bath calculated in the local approach is given by:
\begin{equation}
{\cal J}_h = \omega _h \gamma _h  \left( e^{-\beta _h \omega _h}-\langle a^{\dagger}a\rangle \left(\delta  e^{-\beta _h \omega _h}+1\right)\right)-\frac{\epsilon \gamma _h}{2} \langle X \rangle \left( \delta  e^{-\beta _h \omega _h}+1\right) \nonumber
\end{equation}
placing the steady state solution of Eq.(8) for $\langle a^{\dagger}a\rangle $ and $\langle X \rangle$, we obtain:  
    
\begin{eqnarray}
\begin{array}{ll}
{\cal J}_h = \left( e^{\beta_c \omega_c} - e^{\beta_h\omega_h} \right) \frac{4 \epsilon ^2 \gamma _c \gamma _h e^{\beta _c \omega _c+\beta _h \omega _h} \left(\omega _c \gamma _h e^{\beta _c \omega _c} \left(e^{\beta _h \omega _h}+\delta \right)+\gamma _c \omega _h e^{\beta _h \omega _h} \left(e^{\beta _c \omega _c}+\delta \right)\right)}{\gamma _c^3 \gamma _h e^{2 \beta _h \omega _h} \left(e^{\beta _c \omega _c}+\delta \right){}^3 \left(e^{\beta _h \omega _h}+\delta \right)+2 \gamma _c^2 \left(e^{\beta _c \omega _c}+\delta \right){}^2 e^{\beta _c \omega _c+\beta _h \omega _h}}\cdot\cdot\cdot   \\

~~~~~~~~\frac{}{\times\left(\gamma _h^2 \left(e^{\beta _h \omega _h}+\delta \right){}^2+2 \epsilon ^2 e^{2 \beta _h \omega _h}\right)+\gamma _c \gamma _h e^{2 \beta _c \omega _c} \left(e^{\beta _c \omega _c}+\delta \right) \left(e^{\beta _h \omega _h}+\delta \right) }\cdot\cdot\cdot\\

~~~~~~~~\frac{}{\times\left(4 e^{2 \beta _h \omega _h} \left(\left(\omega _c-\omega _h\right){}^2+2 \epsilon ^2\right)+\gamma _h^2 \left(e^{\beta _h \omega _h}+\delta \right){}^2\right)+4 \epsilon ^2 \gamma _h^2 e^{3 \beta _c \omega _c+\beta _h \omega _h} \left(e^{\beta _h \omega _h}+\delta \right){}^2} \nonumber
\end{array} 
\end{eqnarray}

\subsection{The Global Generator in the Local Representation}
The global approach creates an indirect coupling  of the subsystems with the baths. 
This indirect coupling is evident once we write the the global generator in the local representation, for example, Eq. (11) takes the form:
\begin{eqnarray}
\begin{array}{ll}
{\cal L}_h \rho_s = \gamma_h^+  c^4 \left( \Op a \Op \rho_s \Op a^{\dagger} -\half\{\Op a^{\dagger}\Op a,\Op \rho_s\} 
+ e^{-\beta_h \omega_+}(\Op a^{\dagger} \Op \rho_s \Op a -\half\{\Op a \Op a^{\dagger},\Op \rho_s\})\right)\\

~~~~~~~ +\gamma_h^-  s ^4\left( \Op a \Op \rho_s \Op a^{\dagger} -\half\{\Op a^{\dagger}\Op a,\Op \rho_s\} 
+ e^{-\beta_h \omega_-}(\Op a^{\dagger} \Op \rho_s \Op a -\half\{\Op a \Op a^{\dagger},\Op \rho_s\})\right)\\

~~~~~~~ +\gamma_h^+  c^2s^2\left( \Op b \Op \rho_s \Op b^{\dagger} -\half\{\Op b^{\dagger} \Op b,\rho_s\} 
+ e^{-\beta_h \omega_+}(\Op b^{\dagger} \Op \rho_s \Op  b -\half\{\Op b \Op b^{\dagger},\Op \rho_s\})\right) \\

~~~~~~~ +\gamma_h^-  c^2s^2\left( \Op b \Op \rho_s \Op b^{\dagger} -\half\{\Op b^{\dagger} \Op b,\rho_s\} 
+ e^{-\beta_h \omega_-}(\Op b^{\dagger} \Op \rho_s \Op  b -\half\{\Op b \Op b^{\dagger},\Op \rho_s\})\right) \\

~~~~~~~ +\gamma_h^+ c^3s\left( \Op a \Op \rho_s \Op b^{\dagger}+\Op b \Op \rho_s \Op a^{\dagger} -\half\{\Op a^{\dagger} \Op b + \Op b^{\dagger} \Op a,\rho_s\} 
+ e^{-\beta_h \omega_+}(\Op a^{\dagger} \Op \rho_s \Op  b + \Op b^{\dagger} \Op \rho_s \Op  a -\half\{\Op a^{\dagger} \Op b + \Op b^{\dagger} \Op a,\rho_s\})\right) \\

~~~~~~~ -\gamma_h^- cs^3\left( \Op a \Op \rho_s \Op b^{\dagger}+\Op b \Op \rho_s \Op a^{\dagger} -\half\{\Op a^{\dagger} \Op b + \Op b^{\dagger} \Op a,\rho_s\} 
+ e^{-\beta_h \omega_-}(\Op a^{\dagger} \Op \rho_s \Op  b + \Op b^{\dagger} \Op \rho_s \Op  a -\half\{\Op a^{\dagger} \Op b + \Op b^{\dagger} \Op a,\rho_s\})\right) \\

\end{array} 
\end{eqnarray}
wher we have defined $s \equiv \sin(\theta)$ and $c \equiv \cos(\theta)$.

\subsection{The Covariance Matrix and the Correlation Functions}
We define  a vector of the position and momentum operators $ \xi = (x_A~p_A ~x_B~p_B)$. The covariance matrix is defined through $V_{ij} = \langle \{ \Delta \xi_i, \Delta \xi_j\} \rangle$,
 using the definitions $\{\Delta \xi_i ,\Delta \xi_j \}= \half (\Delta \xi_i \Delta \xi_j +\Delta \xi_j \Delta \xi_i)$ and $\Delta \xi _i = \xi_i -\langle \xi_i \rangle$. The steady state coveraiance matrix is given by\\
\par

$V^{local}$ =
$
\begin{pmatrix}
\langle a^{\dagger}a\rangle +\half & 0 & \half \langle X \rangle & -\half \langle Y \rangle \\
0 & \langle a^{\dagger}a\rangle +\half & \half \langle Y \rangle & \half \langle X \rangle \\
 \half \langle X \rangle & \half \langle Y \rangle & \langle b^{\dagger}b\rangle +\half & 0 \\
 - \half \langle Y \rangle & \half \langle X \rangle & 0 & \langle b^{\dagger}b\rangle +\half
\end{pmatrix}
 $\\
 \\
\par
$V^{global}$ =
$
\left(  \begin{smallmatrix}
\langle d_+^{\dagger}d_+\rangle c^2 +\langle d_-^{\dagger}d_-\rangle s^2 +\half & 0 & (\langle d_+^{\dagger}d_+\rangle  -\langle d_-^{\dagger}d_-\rangle)cs & 0 \\
0 & \langle d_+^{\dagger}d_+\rangle c^2 +\langle d_-^{\dagger}d_-\rangle s^2 +\half & 0 & (\langle d_+^{\dagger}d_+\rangle  -\langle d_-^{\dagger}d_-\rangle)cs  \\
 (\langle d_+^{\dagger}d_+\rangle  -\langle d_-^{\dagger}d_-\rangle)cs  & 0 & \langle d_+^{\dagger}d_+\rangle s^2 +\langle d_-^{\dagger}d_-\rangle c^2 +\half & 0 \\
0 &  (\langle d_+^{\dagger}d_+\rangle  -\langle d_-^{\dagger}d_-\rangle)cs & 0 & \langle d_+^{\dagger}d_+\rangle s^2 +\langle d_-^{\dagger}d_-\rangle c^2 +\half
\end{smallmatrix}\right) 
 $\\
 \\
 with $s \equiv \sin(\theta)$ and $c \equiv \cos(\theta)$. The structure of the covariance matrix in both approaches immediately imply that the two subsystems are separable \cite{simon99}. 
\par
The correlation functions are defined by:
\begin{equation}
cor(\xi_i,\xi_J)=\frac{\langle\Delta\xi_i\Delta\xi_j \rangle}{\sqrt{\langle\Delta\xi_i^2\rangle\langle\Delta\xi_j^2}\rangle}
\end{equation}

\end{document}